\documentclass[12pt]{iopart}
\usepackage{harvard,bm}
\usepackage{iopams,graphicx}   
\newcommand{\rf}[1]{(\ref{#1})}
\begin{document}

\title[]{Sub critical transition to turbulence in three-dimensional Kolmogorov flow}

\author{Lennaert van Veen$^1$\footnote{Corresponding 
author: lennaert.vanveen@uoit.ca} and Susumu Goto$^2$}

\address{$^1$University of Ontario Institute of Technology, 2000 Simcoe Street North, L1H 7K4 Oshawa, Ontario, Canada}
\address{$^2$Graduate School of Engineering Science, Osaka University
1-3 Machikaneyama, Toyonaka, Osaka, 560-8531 Japan}


\begin{abstract}
We study Kolmogorov flow on a three dimensional, periodic domain with aspect ratios fixed to unity. Using an energy method,
we give a concise proof of the linear stability of the laminar flow profile.  Since turbulent motion is observed for high
enough Reynolds numbers, we expect the domain of attraction of the laminar flow to be bounded by the stable manifolds
of simple invariant solutions. We show one such edge state to be an equilibrium with a spatial structure reminiscent of
that found in plane Couette flow, with stream wise rolls on the largest spatial scales. When tracking the edge state, 
we find an upper and a lower branch solution that join in a saddle node bifurcation at finite Reynolds number.
\end{abstract}

\vspace{2pc}
\noindent{\it Keywords}: Kolmogorov flow, box turbulence, sub critical transition, edge states

\maketitle

\section{Introduction}
In his famous 1883 paper, Reynolds investigated the transition from ``direct'' to ``sinuous'' motion of water forced
through a straight smooth tube \cite{Reynolds}. His systematic study lead to the identification of a dimensionless number, now
called the Reynolds number, that essentially measures the ratio of forcing to viscous dissipation. He attempted to use this ratio to
pinpoint the transition -- in current terminology -- from laminar to turbulent flow. His choice of studying pipe flow was
a practical one. Having stated that, given their intractability, integration of the equations of motion was ``not promising'',
he decided to study the transition experimentally in a setup that is relatively easy to realize in a laboratory. Somewhat ironically, 
in choosing ``simplest possible circumstances'', he selected a transition with a highly complicated mathematical structure.
To the best of our current knowledge, the laminar flow in this geometry, known as Hagen-Poiseuille flow, remains linearly stable for
all flow rates. It can be unstable, however, to perturbations with finite amplitude. This type of transition,
which occurs in the absence of a bifurcation of the laminar flow, is now referred to as sub critical. Although Reynolds apparently
expected a linear instability to occur, having attributed to Stokes the observation that this is the general cause for the
onset of sinuous motion, he carefully observed and reported
the benchmarks of sub critical transition: the sinuous motion appears suddenly and with large amplitude and the flow rate
at which it is first observed is very sensitive to the conditions of the in flowing water.

A few decades later, Kolmogorov also considered the transition problem. In a series of seminars in 1958 and 1959,
he proposed to study the problem in a setting more
susceptible to analysis. The seminar series was reported on by \citeasnoun{Arnold}. Conditions at material boundaries, 
highlighted by Reynolds as particularly hard to treat analytically,
are replaced by periodic boundary conditions and instead of three spatial dimensions the problem is posed in two.
A sinusoidal body force is applied on the largest spatial scale to input energy, and this force acts in one direction only.  

In contrast to Reynolds, Kolmogorov explicitly mentions the possibility of a sub critical transition, noting
that experimental work indicates that, with decreasing viscosity, the laminar flows usually become ``either unstable
or so weakly stable, that they are not observed in reality.'' He probably knew both situations to occur in his model,
as results on the linear stability of the laminar flow by \citeasnoun{Meshalkin} are reported in the same paper, albeit without
reference. The result is different for two classes of systems: one in which the domain is elongated in the direction
in which the force acts, and one in which it is elongated in the perpendicular direction. In the former case a linear
bifurcation always takes place, i.e. it is super critical, while in the latter the laminar flow remains linearly stable,
meaning it is sub critical. Kolmogorov hypothesized that in either case, a turbulent solution would exist for small enough
viscosity.

A few years later, \citeasnoun{Iudovich} proved a strong result on the sub critical case. He showed, that in this case
the laminar flow is the only possible limit state, thereby disproving Kolmogorov's hypothesis.
Importantly, the proof includes the square domain. Subsequent studies of Kolmogorov
flow have taken either of two paths: they consider a rectangular domain in the super critical case, or generalize
the model on a square domain, usually by increasing the wave number of the forcing. Work in the former direction has shown, that the
primary bifurcation is a symmetry-breaking steady state bifurcation, while the secondary bifurcation may be of the
Hopf type. Also, it was shown that the locus of the primary
bifurcation approaches infinite Reynolds number as the aspect ratio approaches unity \cite{Okamoto}.
Work on Kolmogorov flow with higher wave number forcing, meanwhile, has lead to a detailed description
of the possible transitions to periodic and, subsequently, turbulent flows \cite{Chen}.

A different, and very natural, generalization of Kolmogorov's model is the inclusion of the third spatial dimension.
Possibly the first attempt to study the three-dimensional case, with both aspect ratios equal to unity, was made
by \citeasnoun{Shebalin}. They discuss the linear stability of the laminar flow in some detail. Citing, as Reynolds did 
more than a century before them, the intractability of this problem, they turn to a conceptual model of ordinary 
differential equations to demonstrate their firm expectation that, for decreasing viscosity, an increasing number
of eigenmodes will become unstable, leading to ``essentially the transition to turbulence described by Landau and
Lifschitz''. In fact, Squire's theorem for parallel shear flows dictates that the addition of the third spatial 
dimension cannot render the laminar flow unstable. On the other hand, their numerical work yields convincing evidence 
that turbulent solutions do exist. This evidence is corroborated by \citeasnoun{Borue}, who used a much higher
spatial resolution, but added hyper viscosity to aid the numerics.

It seems, then, that once again the ``simplest possible circumstances'', being flow on a triply periodic domain
with aspect ratios equal to unity and forcing with the smallest wave number in one direction only, leads to sub critical
transition. In this paper, we commence the study of the transition process through the computation of invariant
solutions, such as equilibria and periodic orbits. In doing so, we hope to strike a compromise between the rigorous
proofs that Kolmogorov had in mind, and ``calculations using computers, which do not completely satisfy a
mathematician.'' \cite{Arnold}.

\section{3D Kolmogorov flow}

The governing equations for viscous, incompressible Kolmogorov flow are
\begin{eqnarray}\label{K-flow}
u_t +\bm{u}\cdot \nabla u + P_x -\nu \Delta u &= \gamma \sin(k_{\rm f}y) \nonumber \\
v_t +\bm{u}\cdot \nabla v + P_y -\nu \Delta v &= 0 \nonumber \\
w_t +\bm{u}\cdot \nabla w + P_z -\nu \Delta w &= 0
\end{eqnarray}
\begin{equation}\label{cont}
\nabla\cdot \bm{u}= 0
\end{equation}
where $\bm{u}=(u,v,w)^t$ is the velocity of a fluid with its constant density fixed to unity, 
$\nu$ the kinematic viscosity and $\gamma$ 
the amplitude of the forcing. We will consider this flow on a cube with linear dimension $L=2\pi$. 
Note, that for quasi-two dimensional solutions with $w\equiv 0$ and
forcing wave number $k_{\rm f}=1$, the system is 
identical to the one posed originally by Kolmogorov \cite{Arnold}.
The boundary conditions are periodic in all directions, but nevertheless we will refer to the $x$ and $z$
directions as {\em stream wise} and {\em span wise}, respectively, in analogy to planar Couette and Poiseuille
flow. We will refer to the $y$ direction as {\em shear wise}.

Mean quantities of special interest are the energy and enstrophy, defined in terms of the velocity and vorticity, $\bm{\omega}=\nabla \times \bm{u}$, as
\begin{eqnarray}\label{eneens}
E&=\frac{1}{L^3}\int \frac{1}{2}|\bm{u}^2|\,\mbox{d}\bm{x} \\
Q&=\frac{1}{L^3}\int \frac{1}{2}|\bm{\omega}^2|\,\mbox{d}\bm{x} 
\end{eqnarray}
as well as the energy input and dissipation rates, given by
\begin{eqnarray}
I&=\frac{\gamma}{L^3}\int u\sin(k_{\rm f}y)\,\mbox{d}\bm{x} \label{eir}\\
D&=2\nu Q \label{edr}
\end{eqnarray}
where all integrals are over the entire periodic domain.

The laminar flow, an exact solution to equations \rf{K-flow}--\rf{cont}, is given by
\begin{equation}\label{laminar}
u^*=\frac{\gamma}{k_{\rm f}^2\nu}\sin(k_{\rm f}y),\ v^*=w^*=0,\ P^*=\mbox{constant}
\end{equation}
and has $E^*=\gamma^2/(4k_{\rm f}^4\nu^2)$, $Q^*=\gamma^2/(4k_{\rm f}^2\nu^2)$ and $I^*=D^*=\gamma^2/(2k_{\rm f}^2\nu)$.

\subsection{Reynolds numbers}

Several Reynolds numbers can be defined for this system. One often used for box turbulence is based on Taylor's
micro scale:
\begin{equation}\label{Re_l}
Re_{\lambda}=\sqrt{\frac{10}{3}}\frac{E}{\nu\sqrt{Q}}
\end{equation}
For the laminar solution, and solutions near it such as described in section \ref{results}, this Reynolds number
scales as $\gamma/\nu^2$, and takes values of the order $10^3-10^4$ for the equilibria presented here. The Taylor
micro scale is supposed to mark the transition from the inertial range to the dissipation range in the
energy spectrum. As will become clear in section \ref{numerical}, the near-laminar solutions do not have
any inertial range in their spectrum, and therefore the Taylor micro scale may not be a useful length
scale to base the Reynolds number on.

A second Reynolds number is based on the energy input rate and was used by \citeasnoun{Linkmann} in their study
of box turbulence with a constant energy input rate:
\begin{equation}\label{Re_3}
Re_{\rm I}=\frac{L^{4/3} I^{1/3}}{\nu}
\end{equation}
This Reynolds number scales as $(\gamma/\nu^2)^{2/3}$ and is of order $10^3$ for the solutions discussed here.

A geometric Reynolds number can be defined using the box length and 
the amplitude of $\bm{u}^*$ as velocity scale:
\begin{equation}\label{Re_2}
Re_{\rm g}=\frac{L\gamma}{k_{\rm f}^2\nu^2}
\end{equation}
For nearly laminar flow, the scaling with $\gamma$ and $\nu$
is the same as that of $Re_{\lambda}$. This definition was used, for instance, by \citeasnoun{Platt} and \citeasnoun{Okamoto}.

The Reynolds number we will use is based on the length scale $1/k_{\rm f}$ and the velocity scale $\sqrt{L\gamma}$ and is given by
$$
Re=\frac{\sqrt{L\gamma}}{k_{\rm f} \nu}=\sqrt{Re_{\rm g}}
$$
In adopting this definition, we follow the direct numerical simulations of \citeasnoun{Shebalin}, as well as 
dynamical systems-based work by \citeasnoun{Chandler}, which is close in spirit to our current study.

\subsection{Stability of the laminar flow}\label{laminar_stab}

Laminar flow profile \rf{laminar} is a parallel shear flow to which Squire's theorem applies. This theorem says
that the first linear instability of the laminar flow, observed when increasing the geometric Reynolds number, must be
independent of the span wise coordinate, i.e. it must be quasi-two dimensional. Combined with the results of 
\citeasnoun{Iudovich}, this result tells us that, for forcing wave number $k_f=1$, there is no linear instability
of the laminar flow.

Since a direct proof is more insightful, we include one based on the energy method, closely following 
\citeasnoun[chapters 3 and 4]{Waleffe}. The starting point is the linearization of equations \rf{K-flow}--\rf{cont}
about the laminar flow
\begin{eqnarray}\label{linearized}
\bm{u}'_t+u^* \bm{u}'_x + Du^* \,v' \bm{e}_x + \nabla P' -\nu \Delta \bm{u}' =0 \nonumber \\
\nabla \cdot \bm{u}'=0
\end{eqnarray}
where $\bm{u}=\bm{u}^*+\bm{u}'$, $P=P^*+P'$, $\bm{e}_x$ is the unit direction vector in the stream wise direction
and $D$ denotes the derivative in the shear wise direction. 
Taking the curl
and selecting the shear wise direction, we find the Squire equation
\begin{equation}\label{Squire}
\left(\partial_t + u^* \partial_x -\nu \Delta\right) \eta + \frac{\mbox{d}u^*}{\mbox{d}y} v'_z =0
\end{equation}
where $\eta$ is the shear wise component of the deviatoric vorticity. Taking the curl twice and selecting the shear wise direction yields the Orr-Sommerfeld equation
\begin{equation}\label{Orr-Sommerfeld}
\left(\partial_t + u^* \partial_x -\nu \Delta\right) \Delta v'- D^2u^* \,v'_x =0
\end{equation}
We write the deviatoric shear wise vorticity and velocity as Fourier modes in the stream wise and span wise directions
\begin{eqnarray}\label{Fourier_expanded}
\eta= \hat{\eta}(y) e^{\lambda t+ i k_x x + i k_z z}\nonumber\\
v'= \hat{v}(y) e^{\lambda t+ i k_x x + i k_z z}
\end{eqnarray}
where $k_x$ and $k_y$ are integers. This yields
\begin{eqnarray}
\left[\lambda+i k_x u^* -\nu(D^2-k_x^2-k_z^2)\right] \hat{\eta}= -i k_z Du^*\,\hat{v}\label{Semispectral1}\\
\left[\lambda+i k_x u^* -\nu \left(D^2-k_x^2-k_z^2\right)\right] \left(D^2-k_x^2-k_z^2\right)\hat{v}=i k_x D^2u^* \,\hat{v}\label{Semispectral2}
\end{eqnarray}
In this equation, the rate of growth of a deviation from laminar flow is the real part of $\lambda$. 
We first filter out four special cases. 
\begin{enumerate}
\item If both $\eta$ and $v'$ are constant in space, the only solutions consistent with the divergence free condition
are of the form $\bm{u}'=(c_1,0,c_2)$ for arbitrary constants $c_{1,2}$. These solutions correspond to Galileo boosts in the stream wise
and span wise directions, which are symmetries of the governing equations \rf{K-flow}. Consequently, these are neutral
modes with $\lambda=0$.
\item If $v'$ is constant in space, but $\eta$ is not, we find from equation \rf{Semispectral1} that $\hat{\eta}\propto \exp(i k_y y)$
and $\lambda=-\nu k_y^2<0$.
\item If $k_x=k_z=0$, but $v'$ depends nontrivially on the shear wise coordinate, we find from equations \rf{Semispectral1} and \rf{Semispectral2} 
that $\hat{\eta},\hat{v}\propto\exp(i k_y y)$
and $\lambda=-\nu k_y^2<0$. 
\item If $k_x^2+k_y^2>0$, but $v'$ does not depend on the shear wise coordinate, equation \rf{Semispectral2} reduces to an algebraic 
equation that admits the solution $\lambda=-\nu (k_x^2+k_z^2)<0$
if and only if $(k_x^2+k_z^2)=k_{\rm f}^2$. 
\end{enumerate}
In deriving the last result, we have used the fact that $D^2u^*=-k_{\rm f}^2 u^*$, which is also pivotal for the
treatment of the more general case below, in which we can assume that $(k_x^2+k_z^2)> 0$ and $\hat{v}$ depends nontrivially on $y$.

We find an energy equation by multiplying equation \rf{Semispectral2} by $\bar{\hat{v}}$, taking the real part and integrating over
the shear wise direction:
\begin{eqnarray}\label{energy}
\fl \Re(\lambda) \int_{y=0}^{2\pi} \left(|D\hat{v}|^2+(k_x^2+k_z^2)|\hat{v}|^2\right)\,\mbox{d}y = \frac{i k_x}{2}\int_{y=0}^{2\pi} Du^*\left(\hat{v}D\bar{\hat{v}}-\bar{\hat{v}}D\hat{v}\right)\,\mbox{d}y\nonumber\\
 -\nu \int_{y=0}^{2\pi} |\left(D^2-k_x^2-k_z^2\right)\hat{v}|^2\,\mbox{d}y
\end{eqnarray}
An enstrophy equation is found by multiplying equation \rf{Semispectral2} by $(D^2-k_x^2-k_y^2)\bar{\hat{v}}$, taking the real part and integrating over
the shear wise direction, which yields
\begin{eqnarray}\label{enstrophy}
\fl \Re(\lambda) \int_{y=0}^{2\pi}|\left(D^2-k_x^2-k_z^2\right)\hat{v}|^2\,\mbox{d}y=-\frac{i k_x}{2}\int_{y=0}^{2\pi} D^3u^*\left(\hat{v}D\bar{\hat{v}}-\bar{\hat{v}}D\hat{v}\right)\,\mbox{d}y\nonumber\\
-\nu \int_{y=0}^{2\pi} \left(|D(D^2-k_x^2-k_z^2)\hat{v}|^2+(k_x^2+k_z^2)|(D^2-k_x^2-k_z^2)\hat{v}|^2\right)\,\mbox{d}y 
\end{eqnarray}
Finally, we eliminate the production term between equations \rf{energy} and \rf{enstrophy}, which results in
\begin{eqnarray}\label{stability}
\fl \Re(\lambda) \int_{y=0}^{2\pi}\left[ \left(2k_x^2+2k_z^2-k_{\rm f}^2 \right) |D\hat{v}|^2+\left(k_x^2+k_z^2-k_{\rm f}^2\right) (k_x^2+k_z^2) |\hat{v}|^2 +|D^2\hat{v}|^2\right]\,\mbox{d}y=\nonumber\\
\!\!\!-\nu\int_{y=0}^{2\pi} \left[\left(k_x^2+k_z^2-k_{\rm f}^2\right) |(D^2-k_x^2-k_z^2)\hat{v}|^2+|D(D^2-k_x^2-k_z^2)\hat{v}|^2\right]\,\mbox{d}y
\end{eqnarray}
For perturbations with $k_x^2+k_z^2\ge k_{\rm f}^2\ge 1$ both integrals are positive and
the growth rate $\Re(\lambda)$ is negative. 



In conclusion we can say that, with the exception of boosts in the stream wise and span wise directions, only deviations from the laminar flow with 
$0<k_x^2+k_z^2< k_{\rm f}$ can have a positive growth rate. For the classical value $k_{\rm f}=1$, this means all deviations are damped.

\subsection{Spatial symmetries}\label{symmetries}

In addition to the Galileo boosts mentioned above, the governing equations with $k_{\rm f}=1$ are equivariant under a large group of symmetries, generated by
\begin{itemize}
\item $T_x(\delta)$ and $T_z(\delta)$, shifts over any distance $\delta$ in the stream wise and span wise directions;
\item $S_z$, reflection in the span wise direction;
\item $S_{xy}$, simultaneous reflection in the stream wise and shear wise directions;
\item $S_{y}$, reflection in the shear wise direction about the plane $y=L/4$;
\item $S_x$, a shift over $L/2$ along the shear wise direction followed by a reflection in the stream wise direction.
\end{itemize}

\section{Numerical simulation and equilibrium solving}\label{numerical}

The simulation code solves for the Fourier coefficients of vorticity, that satisfy
\begin{equation}\label{Four_vort}
\frac{\mbox{d}\widehat{\bm{\omega}}_{\bm{k}}}{\mbox{d} t} = -\bm{k}\times\left(\bm{u}\cdot\nabla \bm{u}\right)\raisebox{3pt}{$\!\widehat{\mbox{}}$}
-\nu \|\bm{k}\|^2 \widehat{\bm{\omega}}-\gamma\cos(y)\bm{e}_z
\end{equation}
where we have used the semi discrete Fourier transform
\begin{equation}\label{SDFT}
\bm{\omega}=\sum\limits_{\bm{k}}\widehat{\bm{\omega}}_{\bm{k}} e^{i \bm{k}\cdot \bm{x}}
\end{equation}
We use the continuity equation, $\bm{k}\cdot \widehat{\bm{u}}_{\bm{k}}=0$, to select two scalar fields to solve for.
The nonlinear term is computed by a standard, FFT-based spectral method and aliasing interactions are removed by the
phase-shift method of \citeasnoun{Patterson}. The spatial resolutions is fixed to $64^3$ grid points and the highest 
wave number Fourier component computed is $k_{\rm max}=30$. In Fig. \ref{shoot_and_spectra} typical three-dimensional
energy spectra are shown for nearly laminar and turbulent flow. The former is clearly well-resolved, while the resolution
is marginal for the latter. Note, that the spectrum of turbulence flow we find is very similar to that of \citeasnoun{Shebalin}.
The dynamical system resulting from this spectral truncation has $N=230,240$ real-valued degrees of 
freedom. The FFTs are computed in parallel, and nearly linear scaling can be achieved up to 64 cores. 
The system is time-stepped with a fourth-order accurate explicit Runge-Kutta-Gill scheme with a fixed time step 
of $1.8\times 10^{-3}$ in units $T=1/\sqrt{k_{\rm f}^2 L\gamma}$, the time scale that $Re$ is based on. The time it
takes to advance the system over one unit $T$ is typically about $30$ CPU minutes on modern processors.

The computation of edge states proceeds in three steps. Each of these steps, and the algorithms used, have become
fairly standard in the study of invariant solutions to the Navier-Stokes equation over the past three decades.
Therefore, we will not discuss them in detail, but rather mention the main points and key papers.

The first step is to identify a long-lived turbulent 
state. We first computed a turbulent state at a high Reynolds number, and then gradually increased the viscosity,
allowing the flow to equilibrate in every step. Around $Re=100$ we start to observe a rapid laminarisation
of the flow, while at $Re=170$, the turbulence is sustained for a few hundred units $T$.

In the second step, we fix the viscosity to the latter value and then bisect between the turbulence state and the laminar state,
following the recipe of \citeasnoun{Itano}. This process is illustrated in Fig. \ref{shoot_and_spectra}(left).
We pick initial conditions on a line in phase space connecting the turbulent to the laminar state, and find
a critical value of the bisection parameter for which the fluid lingers on the boundary of the domain of attraction
of the latter. 
\begin{figure}\label{shoot_and_spectra}
\includegraphics[width=0.45\textwidth]{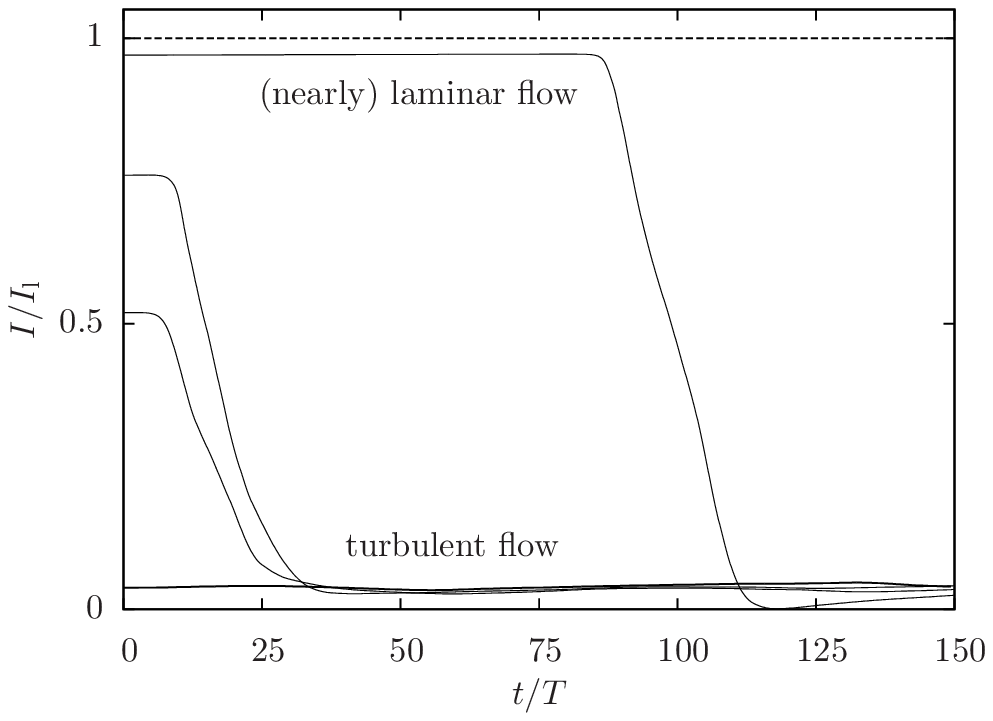}\quad\includegraphics[width=0.4\textwidth]{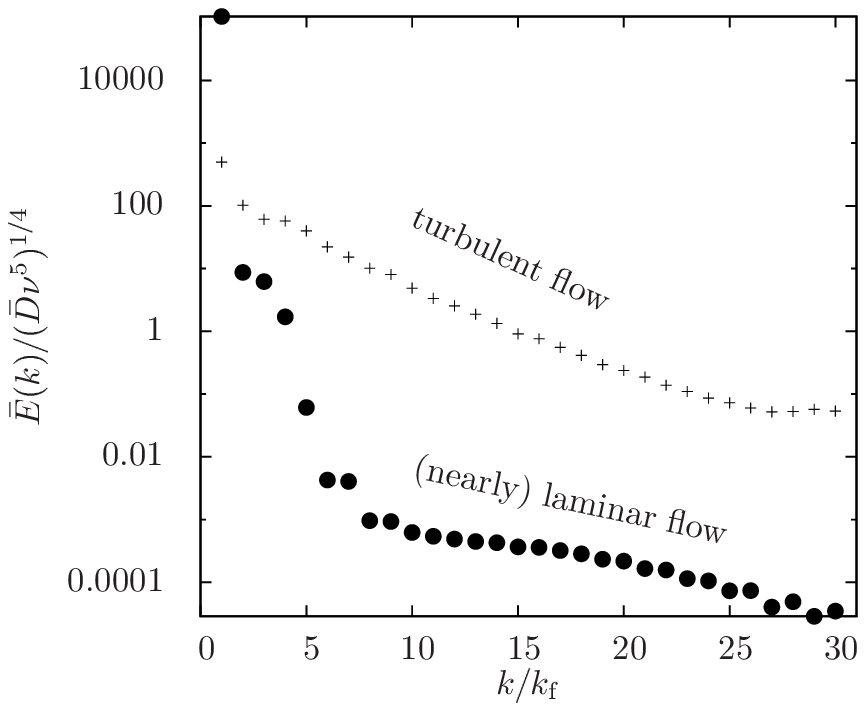}
\caption{Left: Example of bisection to locate the boundary of the domain of attraction of the laminar flow. Shown is the energy input rate,
normalized by its value in laminar flow. The dashed line corresponds to the laminar equilibrium value. Right: three-dimensional energy spectrum of near laminar and turbulent behaviour. The over bars denote temporal means and $D$ is the energy dissipation rate defined in equation \rf{edr}. }
\end{figure}

In the third step, we inspect the behaviour on the boundary and identify transient approaches to
 equilibrium states. Such transient approaches are then used as initial guesses for Newton iteration.
We denote the discretised and spectrally truncated vorticity equation by $\dot{X}=f(X,\nu)$ for a vector $X\in \mathbb{R}^N$ of
unknowns, each corresponding to the real or imaginary part of one of the Fourier coefficients of
vorticity in equation \rf{Four_vort}. The solution of this system of ordinary differential equations
is denoted by $\phi(X,t,\nu)$. We then look for a solution to
\begin{equation}\label{fixed_point}
\phi(X,p,\nu)-T_x(\delta_x)T_z(\delta_z) X =0
\end{equation}
The action of the operators of translation in the stream wise and span wise directions on the vector of unknowns
is readily found from the Fourier transform \rf{SDFT}. A solution to this system of equation is a travelling
wave with speeds $\delta_x/p$ and $\delta_z/p$ in the stream wise and span wise directions, respectively.
Since the number of unknowns is $N+3$, corresponding to the elements of $X$, $\delta_x$ and $\delta_z$ and $\nu$,
we need to add three equations in order to find an isolated solution. Assuming we have a known solution 
$(X_j,\delta_x^j,\delta_z^j,\nu_j)$
and an initial guess for the next solution, $(X^{(0)},\delta_x^{(0)},\delta_z^{(0)},\nu^{(0)})$, we impose that
\begin{eqnarray}\label{phase_conditions}
(X-X^{(0)})\cdot \left.\frac{\mbox{d}T_x(\delta) X^{(0)}}{\mbox{d}\delta}\right|_{\delta=0} + \delta_x-\delta_x^{(0)}=0 \nonumber \\
(X-X^{(0)})\cdot \left.\frac{\mbox{d}T_z(\delta) X^{(0)}}{\mbox{d}\delta}\right|_{\delta=0} + \delta_z-\delta_z^{(0)}=0 \\
(X-X^{(0)})\cdot (X^{(0)}-X_j) +(\delta_x -\delta_x^{(0)})(\delta_x^{(0)}-\delta_x^j)+(\delta_z -\delta_z^{(0)})(\delta_z^{(0)}-\delta_z^j) \nonumber\\
\mbox{\hspace{234pt}} +(\nu-\nu^{(0)}) (\nu^{(0)}-\nu_j) =0 \nonumber
\end{eqnarray}
The first two conditions are that the direction in which we search for the new solution is perpendicular to
the generators of translations in the stream wise and span wise directions. The last condition is that the search
direction is perpendicular to the tangent to the curve of solutions.
The first time we use an initial guess obtained from bisection, no previously computed solution is available,
and we replace the last phase condition by $\nu-\nu^{(0)}=0$, i.e. we keep the viscosity constant.

The resulting system of $N+3$ nonlinear, coupled equations is solved by means of Newton iteration. The linear
system that must be solved for each Newton update step is, in turn, solved by a Krylov subspace method. The resulting
combination of iterative methods is referred to as Newton-Krylov iteration and was first introduced by \citeasnoun{Sanchez}.
If the initial guess is far from the solution, as will generally be the case if it has been filtered from a
time series, the Newton iterates may not converge. In that case, we employ the Newton-hook step, which greatly increases
the radius of convergence, at the cost of increasing the number of necessary iterations. In implementing the Newton-hook
step in conjunction with the Krylov subspace method, we follow \citeasnoun{Viswanath}. This paper also contains
a description of the phase constraints related to the translational symmetries. 

Once the residual of the system of equations \rf{fixed_point}-\rf{phase_conditions}, normalised by the
norm of the solution vector $(X,\delta_x,\delta_z,\nu)$, has dropped below $10^{-5}$, we accept the new point on the solution curve, i.e. we
set $(X_{j+1},\delta_x^{j+1},\delta_z^{j+1},\nu_{j+1})\leftarrow (X,\delta_x,\delta_z,\nu)$. A new initial guess is then generated from known solutions by parameter continuation, 
i.e. changing only $\nu$, or by extrapolation, and the process of solving the nonlinear system 
is repeated. For each equilibrium solution, we compute the four eigenvalues with the largest
real part by means of Arnoldi iteration (see, e.g. \citeasnoun{Gollub}, chapter 9), which is also based on Krylov sub space
iteration and uses the same ingredients as the Newton-Krylov iteration. This algorithm is stopped when
the relative convergence of the eigenvalues is below $10^{-4}$.

\citeasnoun{Sanchez} explain the convergence of the Krylov subspace iteration, which depends on the
integration time $p$. A longer integration time generally leads to faster convergence. On the other hand, the
CPU time taken by each Krylov iteration, which involves time-stepping the vorticity equation and its
linearization, increases linearly with $p$. In this study, we fixed $p=3.5T$. The number of Krylov subspace iterations
necessary to maintain quadratic convergence of the Newton iteration varies from about $30$, on the top branch in figure 
\ref{continuation}, to about $400$ on the bottom branch. Since each equilibrium takes two to four Newton iterations to converge, 
the computation time for the points in this diagram varies from about $40$ to $500$ CPU hours, and the total computation time is 
about one CPU year excluding the eigenvalue computation, which adds another CPU month.

\section{An edge state\label{results}}

Starting from the time series shown in figure \rf{shoot_and_spectra}, the Newton-Krylov-hook method reveals the presence
of an equilibrium state close to the laminar flow. It has no drift in the stream wise or span wise direction, and
has only a single unstable eigenvalue. 
A continuation
of this state in the Reynolds number is shown in figure \ref{continuation}. As the Reynolds number increases, its bulk 
properties, such as the energy input rate plotted in the figure, approach that of the laminar flow. Following the curve
in the other direction, we find a saddle-node bifurcation around $Re=142.5$. The equilibria on the bottom branch are
completely stable up to $Re=145.3$, where a Hopf bifurcation occurs. Beyond this point, several more Hopf bifurcations
are crossed and the number of unstable eigenvalues increases rapidly. 
\begin{figure}\label{continuation}
\begin{center}
\includegraphics[width=0.55\textwidth]{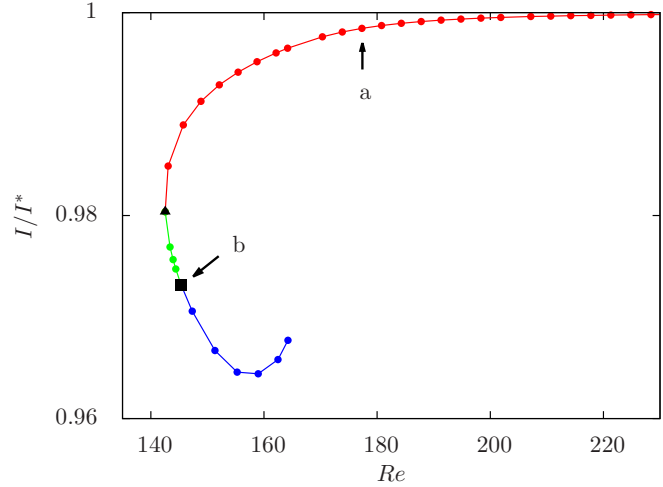}
\end{center}
\caption{Continuation of the equilibrium state in the Reynolds number. On the vertical axis, the energy input rate is shown, normalized by
its value in laminar flow. Linear stability is indicated by colour as follows: edge states, with a single positive eigenvalue,
are shown in red, stable states in green and states with at least two unstable eigenvalues in blue. The solid dots indicate the
solutions computed during the continuation. The solid triangle denotes a saddle-node bifurcation and the solid square a Hopf 
bifurcation. Labels a and b correspond to the physical space portraits in figure \ref{portraits}.}
\end{figure}

\begin{figure}\label{portraits}
\includegraphics[width=0.45\textwidth]{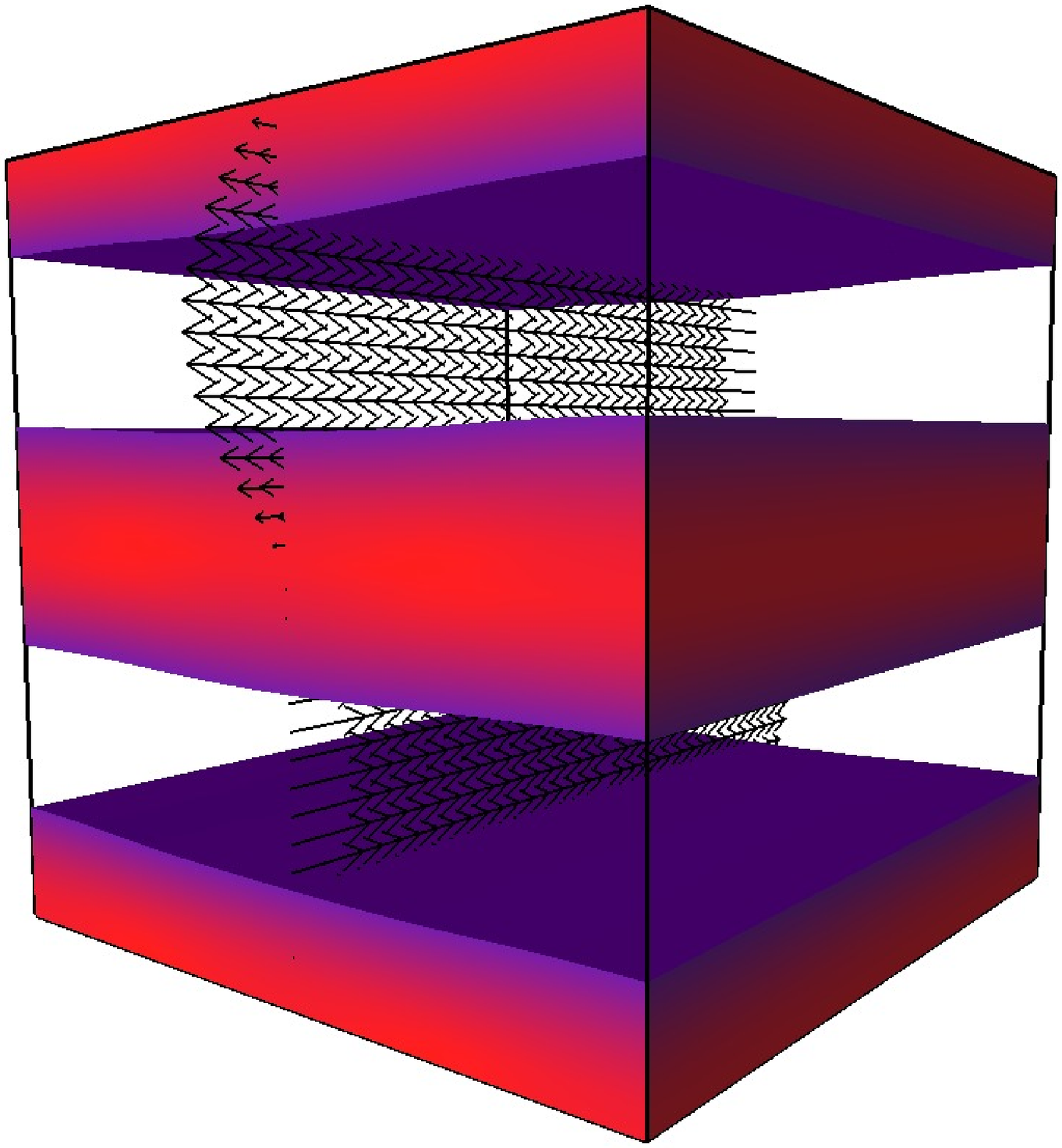}\quad\includegraphics[width=0.46\textwidth]{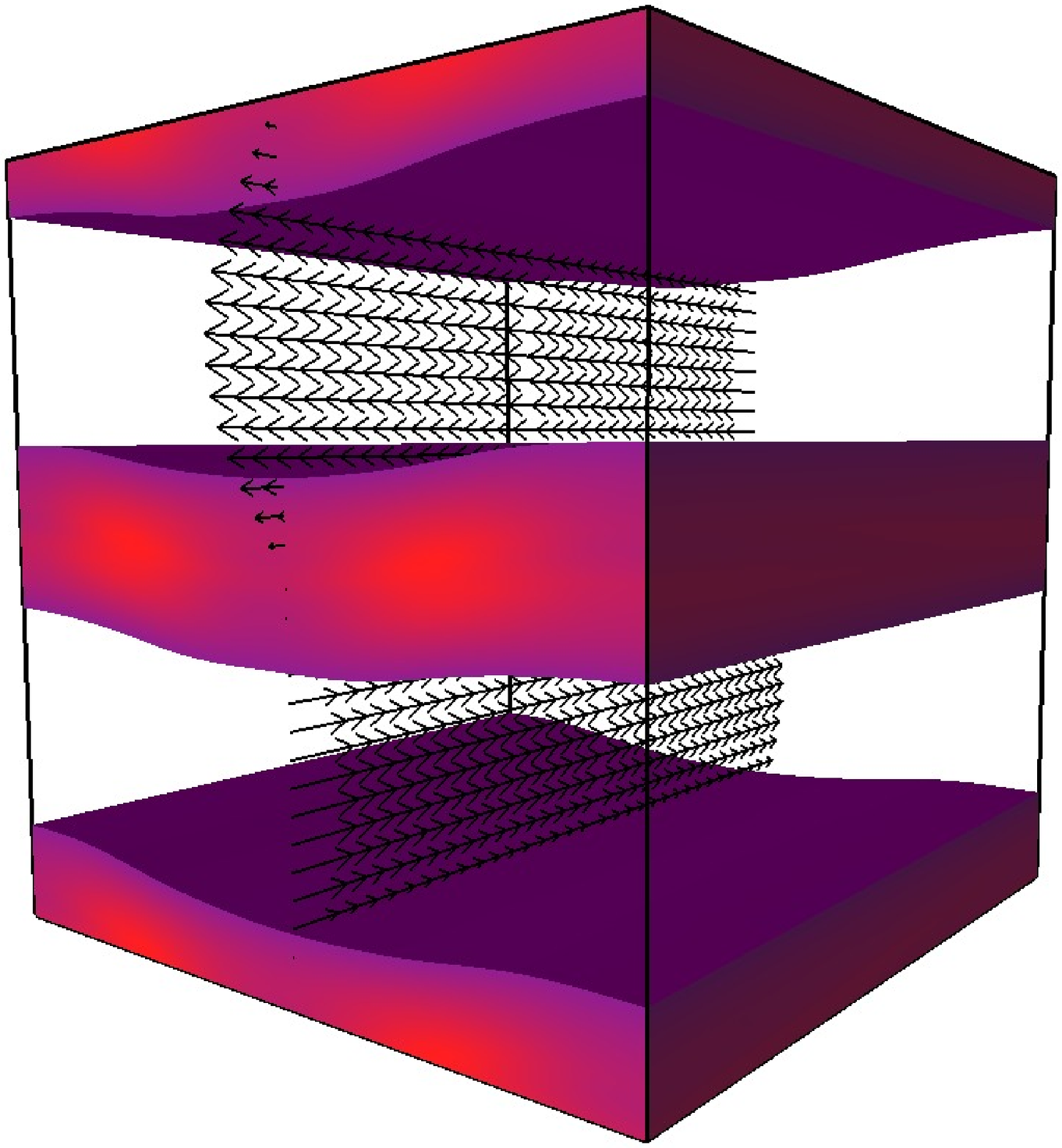}\\
\begin{picture}(0,0)
\put(17,5){(a)}
\put(236,5){(b)}
\put(210,15){\vector(3,2){13}}
\put(210,15){\vector(-2,1){13}}
\put(210,15){\vector(0,1){19}}
\put(225,23){$x$}
\put(208,40){$y$}
\put(190,20){$z$}
\end{picture}
\caption{(a): Physical space portrait of an edge state in figure \ref{continuation}. Shown are the velocity in the plane
$z=\pi$ and isolevels of the vorticity ranging from red (high) to blue (low). The lowest isolevel shown is at 50\%
of the maximal value. (b): As in (a) for the equilibrium at the Hopf bifurcation in figure \ref{continuation}.}
\end{figure}

Two physical space visualizations of equilibrium solutions are shown in figure \ref{portraits}. Their structure
is similar to that of some nearly laminar equilibria found in channel flow. On top of the stream wise and span wise independent
shear flow, induced by the body force, stream wise vortices are formed and the isosurfaces of vorticity exhibit
a weak, sinusoidal variation in the span wise direction.

It can be confirmed by an inspection of the Fourier coefficients of vorticity that this family of equilibria is,
up to round-off and truncation error, invariant under symmetries $S_x$, $S_z$ and $T_z(L/2)\circ S_{xy}$.
The latter symmetry is found to hold approximately in turbulent minimal plane Couette flow, and is sometimes
imposed on that flow from the outset, as by \citeasnoun{Kawahara}. A second symmetry they imposed is
$T_x(L/2)\circ S_z$. While our equilibrium only depends weakly on the stream wise direction, it does not
appear to have any discrete translational symmetry in this direction.


\section{Discussion}

The observation that Kolmogorov flow on a three-dimensional domain with aspect ratios fixed to unity follows a
sub critical transition to turbulence is perhaps surprising in the light of previous results on spatially
periodic flows. The laminar state in ABC flow, for instance, was shown to turn unstable in a super critical Hopf
bifurcation \cite{Ashwin} and the same instability occurs in Kida-Pelz flow \cite{Kida,vanVeen}. Similarly,
the primary instability of laminar Kolmogorov flow with a forcing wave number greater than one will be that
of the corresponding flow in two dimensions, either pitchfork of Hopf.

On the other hand, recent results by \citeasnoun{Linkmann} indicate that a forcing mechanism often used in
large-scale simulations of homogeneous isotropic turbulence may also generate turbulence in the presence
of a stable laminar flow. This forcing mechanism keeps the energy input rate constant in time by scaling a
number of low wave number Fourier components in every time step. Because of the difference in the nature of
the forcing, a direct comparison is difficult, but it seems their results have a much lower Reynolds number
than those presented here. Their Reynolds number based on the energy input rate, $R_{I}$, varies in the range $54$ -- $98$
while in our flow it takes values from about $Re_{I}=800$, in turbulent flow, to $Re_I=2500$, in near laminar flow. 
Another qualitative difference is the fact that the laminar flow found in their system depends on the 
initial condition as the forcing depends on the instantaneous flow field.

We hope, that a further study of the edge state will shed new light on Kolmogorov's original question about
the nature of laminar to turbulent transition in spatially periodic flows. We are currently computing time-periodic 
solutions, such as those branching off the current equilibrium in various Hopf bifurcations. The dynamics of
-- possibly transient -- turbulence can be investigated by computing periodic solutions far removed from
the laminar state, along the lines of \citeasnoun{Chandler}. In addition, it will be interesting to compare
the results to those obtained in minimal plane Couette flow to see what aspects of the dynamics of transition
are directly related to the presence of material boundaries. Finally, we expect that three-dimensional
Kolmogorov flow will prove a fertile test ground for new algorithms for the computation of invariant
solutions since the absence of material boundaries, and the associated polynomial basis functions,
renders the simulation algorithm comparatively simple.

\ack

Author LvV was supported by long-term visiting scholarship L14708 of the Japan Society for Promotion of Science during his sabbatical stay at
Osaka University, where this work was started. We thank Fedor Naumkin for his translation of \citeasnoun{Arnold}.

\section*{References}
\bibliographystyle{jphysicsB}
\bibliography{subkol}

\end{document}